\def\strutdepth{\dp\strutbox}
\def\ergoout#1{\vtop to \strutdepth%
 {\baselineskip\strutdepth\vss\llap{\small #1}\null}}
\def\Ergomargin#1{\strut\vadjust{\kern-\strutdepth\ergoout#1}}
\expandafter\chardef\csname pre EtBbb at\endcsname=\the\catcode`\@
\catcode`\@=11
\font\elvmsb=msbm10\@halfmag\font\egtmsb=msbm8\font\sixmsb=msbm6
\newfam\msbfam
\textfont\msbfam=\elvmsb
\scriptfont\msbfam=\egtmsb
\scriptscriptfont\msbfam=\sixmsb
\def\ErgoBbb#1{\fam\msbfam\relax#1}
\catcode`\@=\csname pre EtBbb at\endcsname
\documentstyle[11pt]{article}
\begin{document} 
\title{Calculation of the $\varphi ^{4}$ 6-loop non-zeta transcendental} 
\author{Oliver Schnetz 
\thanks{Institut f{\"u}r theoretische Physik III, Staudtstra{\ss}e 
7, 91058 Erlangen, Germany\newline 
\hspace*{4ex}e-mail: schnetz@theorie3.physik.uni-erlangen.de}} 
\date{December 16, 1999} 
\maketitle 
\begin{abstract} 
We present an analytic calculation of the first transcendental in 
$\varphi ^{4}$-Theory that is not of the form $ \zeta (2n+1)$. It 
is encountered at 6 loops and known to be a weight 8 double sum. 
Here it is obtained by reducing multiple zeta values of depth $ 
\le 4$. We give a closed expression in terms of a zeta-related sum 
for a family of diagrams that entails a class of physical graphs. 
We confirm that this class produces multiple zeta values of weights 
equal to the crossing numbers of the related knots. 
\end{abstract} 
\tableofcontents 
\section{Introduction} 
The miraculous connection between Feynman graphs, knots, and numbers 
discovered by D. Kreimer and D.J. Broadhurst \cite{a}--\cite{e} 
reveals a structure underlying generic renormalizable Quantum Field 
Theories. It is conjectured that via Knot Theory the topology of 
an amplitude provides its transcendental number contents which for 
primitively divergent diagrams, is given by the renormalization 
scheme independent numerical factor in front of the logarithmic 
singularity.

The calculation of these numbers was acchieved in \cite{a} for 56 
of the 59 primitively divergent graphs up to 7 loops in massless 
$\varphi ^{4}$-Theory. Diagrams that could not be calculated analytically 
were evaluated numerically to high precision. A systematic search 
for the rational coefficients of a conjectured basis of transcendentals 
led in all but three 7-loop diagrams (related to the hyperbolic 
knots 10$_{139}$ and 10$_{152}$) to (with very high probability) 
exact results.

The first graphs where analytic calculations were not available showed 
up at 6 loops. These diagrams entailed the 4,3-torus knot (as opposed 
to the $ (2n+1)$,2-torus knots leading to $ \zeta (2n+1)$) which 
was determined to be related to the transcendental $ \frac{1}{5
}(-\zeta (5,3)+29\zeta (8))$, where $ \zeta (n,m)=\sum _{k>
\ell \ge 1}k^{-n}\ell ^{-m}$. In the terminology of \cite{a} these 
graphs are the $ M(2,1,1,0)$ and the $ M(1,1,1,1)$ '4-banana' diagrams.

In this paper we focus on the more symmetric $ M(1,1,1,1)$. We keep 
equations general as far as this does not complicate calculations. 
In Theorem 10 a closed result is given that easily generalizes to 
encorporate all $n$-banana diagrams (defined below). The derivation 
is organized as a collection of propositions that are all proved 
in full detail although most of them are elementary.

\section{Notation and definitions}

We normalize 4-dimensional integrals with $ \frac{1}{4\pi ^{2}}$ 
and 2-dimensional integrals with $ \frac{1}{2\pi }$ (which is symmetric 
under Fourier transformation), 
\begin{equation}
\int dx\equiv \frac{1}{4\pi ^{2}{}} \int _{{\ErgoBbb R}^{4}}d^{4
}x\hspace{.6ex},\hspace{2ex}\int _{{\ErgoBbb C}}dx\equiv \frac{
1}{2\pi } \int _{{\ErgoBbb R}^{2}}d^{2}x\hspace{.6ex}.
\end{equation}

In order to handle the general case most efficiently, it is convenient 
to use a $\delta $-symbol (also $ \bar{\delta }$, $\delta _{i}$, 
etc.) with the meaning that $ f(\delta )$ is the constant term in 
the Laurant expansion of $f$ at $ \delta =0$, 
\begin{equation}
f\left( \delta \right) \equiv f_{0}=\frac{1}{N!} \left. \frac{
\partial ^N}{\partial x^N}\right| _{x=0}x^Nf\left( x\right) \hbox{ 
, if }f\left( x\right) =\sum _{n=-N}^{\infty }f_{n}x^{n}
\hspace{.6ex}.
\end{equation} 
Expample: 
\begin{displaymath}
\frac{1}{\delta ^{k}{}} x^\delta =\frac{1}{k!} \hbox
{\hspace{.38ex}ln\hspace{.38ex}}^{k}\left( x\right) \hbox{ , if }k
\ge 0\hbox{ (and zero otherwise).\hspace{.38ex}}
\end{displaymath}

\pagebreak[3]

\noindent {\bf Definition} 1.\ For $ x,y\in {\ErgoBbb R}^{4}$ let 
$ f_{k}(x,y)$ be recursively defined by 
\begin{eqnarray}
f_{0}\left( x,y\right) &=&\frac{1}{\left( x-y\right) ^{2}{}}\\ 
f_{k+1}\left( x,y\right) &=&\int dz\frac{1}{\left( x-z\right) ^{
2}{}} \frac{1}{z^{2}{}} f_{k}\left( z,y\right) 
\end{eqnarray} 
\pagebreak[3]

\noindent {\bf Remark} 2.\ The funcion $ f_{k}(x,y)$ depends only 
on $ |x|$, $ |y|$, and the angle between $x$ and $y$. We may treat 
it as a function of two complex variables $ x_{\ErgoBbb C}$ and 
$ y_{\ErgoBbb C}$ that coincide with $x$ and $y$ if the complex 
plane is considered as the subspace of ${\ErgoBbb R}^{4}$ spanned 
by $x$ and $y$.

Instead of $ f_{k}(x_{\ErgoBbb C},1_{\ErgoBbb C})$ we also write 
$ f_{k}(x,\bar{x})$. We have 
\begin{equation}
\label{1}f_{k}\left( x,y\right) =f_{k}\left( y,x\right) 
\hspace{.6ex},\hspace{2ex}f_{k}\left( x_{\ErgoBbb C},y_
{\ErgoBbb C}\right) =\frac{1}{|y_{\ErgoBbb C}|^{2}{}} f\left( x_
{\ErgoBbb C}/y_{\ErgoBbb C},1_{\ErgoBbb C}\right) =\frac{1}{|x_
{\ErgoBbb C}|^{2}{}} f\left( y_{\ErgoBbb C}/x_{\ErgoBbb C},1_
{\ErgoBbb C}\right) \hspace{.6ex}.
\end{equation} 
\pagebreak[3]

\noindent {\bf Definition} 3 ($n$-banana). Let $ {\bf k}=(k_{1
},{\ldots},k_{n})$ be a vector of non-negative integers with not 
more than one of the $ k_{i}=0$. Let $ -2<\alpha <n-2$ and $ e_{
1}$ be a 4-dimensional unit-vector. Then 
\begin{equation}
B_\alpha \left( {\bf k}\right) =\int dx\left( x^{2}\right) ^
\alpha \prod _{i=1}^{n}f_{k_{i}}\left( x,e_{1}\right) 
\hspace{.6ex}.
\end{equation} 
Note, that the integral is independent of $ e_{1}$ and converges 
under the above conditions.

In the notation of \cite{a} we have 
\begin{displaymath}
G\left( k_{1},k_{2},k_{3}\right) =B_{0}\left( k_{1},k_{2},k_{3}
\right) =B_{-1}\left( k_{1},k_{2},k_{3}\right) \hbox{ and }M
\left( k_{1},k_{2},k_{3},k_{4}\right) =B_{0}\left( k_{1},k_{2},k
_{3},k_{4}\right) \hspace{.6ex}.
\end{displaymath}

\section{Calculation of the $n$-banana diagram} 
We restrict ourselves to $ -1<\alpha <n-3$ and $ k_{i}\ge 1
\hspace*{1ex}\forall i$ which includes $ B_{0}(1,1,1,1)$.

\pagebreak[3]

\noindent {\bf Proposition} 4.\  
Let $ f:{\ErgoBbb R}^{4}\rightarrow {\ErgoBbb R}$, $ x\mapsto f(x
)$ be a function of $ |x|$ and $ x\cdot e_{1}$ only, then 
\begin{equation}
\int dxf\left( x\right) =-\frac{1}{4}\int _{{\ErgoBbb C}}dx
\left( x-\bar{x}\right) ^{2}f_{\ErgoBbb C}\left( x,\bar{x}
\right) \hspace{.6ex},
\end{equation} 
where $ f_{\ErgoBbb C}:{\ErgoBbb C}\rightarrow {\ErgoBbb R}$, $ x
\mapsto f_{\ErgoBbb C}(x,\bar{x})$ equals $f$ on the complex plane 
spanned by $x$ and $ e_{1}=1_{\ErgoBbb C}$.

\pagebreak[3]

\noindent {\bf Proof}. 
Introducing polar coordinates yields 
\begin{eqnarray*}
&&\int dxf\left( x\right) \hspace*{1ex}=\hspace*{1ex}\frac{1}{4
\pi ^{2}{}} \int _{0}^{\infty }dx x^{3}4\pi \int _{0}^{\pi }d
\vartheta \sin^{2}\vartheta f_{\ErgoBbb C}\left( xe^{i
\vartheta },xe^{-i\vartheta }\right) \\ 
&=&\frac{1}{2\pi } \int _{0}^{\infty }dx x\int _{0}^{2\pi }d
\vartheta \frac{\left( xe^{i\vartheta }-xe^{-i\vartheta }
\right) ^{2}}{\left( 2i\right) ^{2}{}} f_{\ErgoBbb C}\left( xe^
{i\vartheta },xe^{-i\vartheta }\right) \hspace*{1ex}=
\hspace*{1ex}-\frac{1}{4}\int _{{\ErgoBbb C}}dx\left( x-\bar{x}
\right) ^{2}f_{\ErgoBbb C}\left( x,\bar{x}\right) 
\hspace{.6ex}.
\end{eqnarray*}\hfill $\Box $
\pagebreak[3] 

\noindent{}Hence, 
\begin{displaymath}
B_\alpha \left( {\bf k}\right) =-\frac{1}{4}\int _{{\ErgoBbb C}}dx
\left( x\bar{x}\right) ^\alpha \left( x-\bar{x}\right) ^{2}
\prod _{i=1}^{n}f_{k_{i}}\left( x,\bar{x}\right) \hspace{.6ex}.
\end{displaymath} 
With Eq.\ (\ref{1}) we immediately obtain 
\begin{equation}
B_\alpha \left( {\bf k}\right) =B_{n-4-\alpha }\left( {\bf k}
\right) \hspace{.6ex}.
\end{equation} 
Moreover, with $ x=|x|\hat{x}$, $ \bar{x}=|x|/\hat{x}$ we get 
\begin{equation}
\label{4}B_\alpha \left( {\bf k}\right) =-\frac{1}{4}\int _{0}^{
\infty }dxx^{2\alpha +1}\int _{\partial B}\frac{d\hat{x}}{2\pi i
\hat{x}} \left( x\hat{x}-x/\hat{x}\right) ^{2}\prod _{i=1}^{n}f_
{k_{i}}\left( x\hat{x},x/\hat{x}\right) \hspace{.6ex},
\end{equation} 
where we integrate $ \hat{x}$ along the unit circle $ \partial B
$ in ${\ErgoBbb C}$.

\pagebreak[3]

\noindent {\bf Proposition} 5.\  
\begin{equation}
\label{3}f_{k}\left( x,\bar{x}\right) =-\frac{\delta + \bar
{\delta }}{\left( 4\delta \bar{\delta }\right) ^{k}{}} \int _{0
}^{1}\!dt\;t^\delta \left( \frac{t}{x\bar{x}}\right) ^{\bar
{\delta }}\frac{1}{\left( t-x\right) } \frac{1}{\left( t-\bar
{x}\right) } \hspace{.6ex}.
\end{equation} 
With (\ref{1}) and our definition of the $\delta $-symbol we get 
the following 4-dimensional statement.

\pagebreak[3]

\noindent {\bf Corollary} 6.\  
\begin{equation}
f_{k}\left( x,y\right) =-\frac{1}{4^{k}k!\left( k-1\right) !} 
\int _{0}^{1}\!dt\hbox{\hspace{.38ex}ln\hspace{.38ex}}^{k-1}
\left( t\right) \hbox{\hspace{.38ex}ln\hspace{.38ex}}^{k-1}
\left( \frac{ty^{2}}{x^{2}}\right) \hbox{\hspace{.38ex}ln}
\left( \frac{t^{2}y^{2}}{x^{2}}\right) \frac{1}{\left( yt-x
\right) ^{2}{}} \hspace{.6ex}.
\end{equation}

\pagebreak[3]

\noindent {\bf Proof} of Prop.\ 5.\  
Expansion of the propagator into a sum of Gegenbauer polynomials 
gives (with $ \cos(\vartheta _{xy})=x\cdot y/|xy|$ and $ x_<{} 
=x$ if $ x<1$, $ x_<{} =1/x$ if $ x>1$) 
\begin{equation}
\frac{|xy|}{\left( x-y\right) ^{2}{}} =\sum _{n=1}^{\infty }C_{n
-1}\left( \cos\left( \vartheta _{xy}\right) \right) \left| 
\frac{x}{y}\right| _<^{n}=\int \frac{dP}{\pi } \sum _{n=1}^{
\infty }nC_{n-1}\left( \cos\left( \vartheta _{xy}\right) 
\right) \left| \frac{x}{y}\right| ^{iP}\frac{1}{\left( n^{2}+P^{
2}\right) } \hspace{.6ex}.
\end{equation} 
Orthogonality of the Gegenbauer Polynomials yields 
\begin{equation}
f_{k}\left( x,e_{1}\right) =\int \frac{dP}{\pi } \sum _{n=1}^{
\infty }nC_{n-1}\left( \cos\left( \vartheta _{xe_{1}}\right) 
\right) |x|^{iP-1}\frac{1}{\left( n^{2}+P^{2}\right) ^{k+1}{}} 
\end{equation} 
from which we derive the recursion relation 
\begin{eqnarray*}
&&\frac{\partial }{\partial |x|} |x|f_{k}\left( x,e_{1}\right) =
\int \frac{dP}{\pi } \sum _{n=1}^{\infty }nC_{n-1}\left( \cos
\left( \vartheta _{xe_{1}}\right) \right) |x|^{iP-1}\frac{-i}{2
k}\frac{\partial }{\partial P} \frac{1}{\left( n^{2}+P^{2}
\right) ^{k}{}}\\ 
&=&\frac{i}{2k}\int \frac{dP}{\pi } \sum _{n=1}^{\infty }nC_{n-1
}\left( \cos\left( \vartheta _{xe_{1}}\right) \right) \frac{1}{
\left( n^{2}+P^{2}\right) ^{k}{}} \frac{\partial |x|^{iP-1}}{
\partial P} =-\frac{\hbox{\hspace{.38ex}ln}\left( |x|\right) }{
2k}f_{k-1}\left( x,e_{1}\right) .
\end{eqnarray*} 
If we integrate this relation we encounter an ambiguity of the form 
$ c(\hat{x})/|x|$ which is removed by the condition $ \lim_{|x| 
\rightarrow \infty }|x|f(x,e_{1})=0$. Equation (\ref{3}) meets this 
condition. Moreover with $ g_{k}(x,\bar{x})=\int _{0}^{1}\!dt\;t^
\delta (t/(x\bar{x}))^{\bar{\delta }}(1/(t-x)-1/(t-\bar{x}))$ we 
get 
\begin{eqnarray*}
&&\hspace{-.5cm} -\frac{2k}{\hbox{\hspace{.38ex}ln}\left( |x|
\right) } \frac{\partial }{\partial |x|} |x|\frac{\delta + \bar
{\delta }}{\left( 4\delta \bar{\delta }\right) ^{k}{}} \frac{1}
{\bar{x}-x} g_{k}\left( x,\bar{x}\right) \\ 
&&\hspace{-.5cm}=-\frac{2k}{\hbox{\hspace{.38ex}ln}\left( |x|
\right) } \frac{\delta + \bar{\delta }}{\left( 4\delta \bar
{\delta }\right) ^{k}{}} \frac{1}{1/\hat{x}-\hat{x}} \left( -
\frac{2\bar{\delta }}{|x|} g_{k}\left( x,\bar{x}\right) +\int _{
0}^{1}\hspace{-1ex} dt\;t^\delta \left( \frac{t}{x\bar{x}}
\right) ^{\bar{\delta }}\left( \frac{\hat{x}}{\left( t-x
\right) ^{2}{}} -\frac{\hat{x}^{-1}}{\left( t-\bar{x}\right) ^{
2}}\right) \right) \\ 
&&\hspace{-.5cm}=-\frac{2k}{\hbox{\hspace{.38ex}ln}\left( |x|
\right) } \frac{\delta + \bar{\delta }}{\left( 4\delta \bar
{\delta }\right) ^{k}{}} \frac{1}{\bar{x}-x} \left( -2\bar
{\delta }g_{k}\left( x,\bar{x}\right) +\left( -\frac{x}{1-x}+
\frac{\bar{x}}{1-\bar{x}}\right) \left( x\bar{x}\right) ^{-\bar
{\delta }}+\left( \delta + \bar{\delta }\right) g_{k}\left( x,
\bar{x}\right) \right) \\ 
&&\hspace{-.5cm}=-\frac{2k}{4^{k}\hbox{\hspace{.38ex}ln}\left( |x
|\right) \left( \bar{x}-x\right) } \int _{0}^{1}dt\left( \frac{\hbox
{\hspace{.38ex}ln\hspace{.38ex}}^{k}\left( t/x\bar{x}\right) }{k!
} \frac{\hbox{\hspace{.38ex}ln\hspace{.38ex}}^{k-2}\left( t
\right) }{\left( k-2\right) !} -\frac{\hbox{\hspace{.38ex}ln\hspace{.38ex}}
^{k-2}\left( t/x\bar{x}\right) }{\left( k-2\right) !} \frac{\hbox
{\hspace{.38ex}ln\hspace{.38ex}}^{k}\left( t\right) }{k!}
\right) \left( \frac{1}{t-x} -\frac{1}{t-\bar{x}}\right) \\ 
&&\hspace{ 1ex}\hbox{ if }k\ge 2\hspace*{4ex}\hbox
{\hspace{.38ex}and\hspace{.38ex}}\hspace*{4ex}\frac{1}{1-x} 
\frac{1}{1-\bar{x}} \hspace*{4ex}\hbox{\hspace{.38ex}if }k=1.
\end{eqnarray*} 
According to Cor.\ 6 this equals $ f_{k-1}(x,\bar{x})$ which completes 
the proof.\hfill $\Box $
\pagebreak[3] 

We find the following behaviour of $ f_{k}(x\hat{x},x/\hat{x})$ as 
a function of a complex variable $ \hat{x}$.

\pagebreak[3]

\noindent {\bf Proposition} 7.\  
The function $ f_{k}(x\hat{x},x/\hat{x})$ is analytic in $ \hat
{x}$ save for two cuts on the real axis ranging from 0 to $ x_<
$ and from $ 1/x_<$ to $\infty $. Moreover, if $ \hat{x}=|\hat
{x}|e^{i\varepsilon }$ we have in the limit $ \varepsilon 
\rightarrow 0$ 
\begin{equation}
f_{k}\left( x\hat{x},x/\hat{x}\right) -f_{k}\left( x\overline {
\hat{x}},x/\overline {\hat{x}}\right) =2\pi i\frac{\delta + 
\bar{\delta }}{\left( 4\delta \bar{\delta }\right) ^{k}{}} 
\frac{\left( x|\hat{x}|\right) ^\delta \left( x/|\hat{x}|
\right) ^{-\bar{\delta }}}{x/|\hat{x}|-x|\hat{x}|} \left( 
\Theta \left( x_<-|\hat{x}|\right) -\Theta \left( |\hat{x}|-1/x_
<\right) \right) \hspace{.6ex}.
\end{equation}

\pagebreak[3]

\noindent {\bf Proof}. 
\begin{eqnarray*}
\hspace{-5mm}&&f_{k}\left( x\hat{x},x/\hat{x}\right) -f_{k}
\left( x\overline {\hat{x}},x/\overline {\hat{x}}\right) \\ 
\hspace{-5mm}&=&-\frac{\delta + \bar{\delta }}{\left( 4\delta 
\bar{\delta }\right) ^{k}{}} \int _{0}^{1}\hspace{-1ex} dt\;t^
\delta \left( \frac{t}{x^{2}}\right) ^{\bar{\delta }}\left( 
\left( \frac{1}{t-x\hat{x}} -\frac{1}{t-x\overline {\hat{x}}}
\right) \frac{1}{t-x/\hat{x}} +\frac{1}{t-x\overline {\hat{x}}} 
\left( \frac{1}{t-x/\hat{x}} -\frac{1}{t-x/\overline {\hat{x}}}
\right) \right) \hspace{.6ex}.
\end{eqnarray*} 
(In the limit $ \varepsilon \rightarrow 0$ the poles in the differences 
move to the points $ t=x|\hat{x}|$ and $ t=x/|\hat{x}|$ on the real 
axis. In the case that one of them falls into the interval [0,1] 
we deform the $t$-integrals to closed curves $ C_{x|\hat{x}|}$ and 
$ C_{x/|\hat{x}|}^{-1}$  encircling the singularities with positive 
or negative orientation.) 
\begin{eqnarray*}
\hspace{-5mm}&=&-\frac{\delta + \bar{\delta }}{\left( 4\delta 
\bar{\delta }\right) ^{k}{}} \left( \Theta \left( 1-x|\hat{x}|
\right) \int _{C_{x|\hat{x}|}}dt+\Theta \left( 1-x/|\hat{x}|
\right) \int _{C^{-1}_{x/|\hat{x}|}}dt\right) t^\delta \left( 
\frac{t}{x^{2}}\right) ^{\bar{\delta }}\left( \frac{1}{t-x|\hat
{x}|} \frac{1}{t-x/|\hat{x}|}\right) \\ 
\hspace{-5mm}&=&2\pi i\frac{\delta + \bar{\delta }}{\left( 4
\delta \bar{\delta }\right) ^{k}{}} \frac{1}{x/|\hat{x}|-x|\hat
{x}|} \left( \Theta \left( 1-x|\hat{x}|\right) \left( x|\hat{x}
|\right) ^\delta \left( x/|\hat{x}|\right) ^{-\bar{\delta }}+
\Theta \left( 1-x/|\hat{x}|\right) \left( x|\hat{x}|\right) ^{-
\bar{\delta }}\left( x/|\hat{x}|\right) ^\delta \right) .
\end{eqnarray*} 
Changing $\delta $ to $ -\bar{\delta }$ and $ \bar{\delta }$ to $ 
-\delta $ in the second term leads to the claimed result.\hfill $\Box $
\pagebreak[3] 

\noindent{}Since, for $ n\ge 3$ the integrand in Eq.\ (\ref{4}) has 
no pole at $ \hat{x}=0$ we may deform the integration contour to 
range from 0 to $ e^{-i\varepsilon }$ and back from $ e^{i
\varepsilon }$ to 0 (in the limit $ \varepsilon \rightarrow 0$). 
\begin{eqnarray}
\hspace{-5mm}&&\int _{\partial B}\frac{d\hat{x}}{2\pi i\hat{x}} 
\left( x\hat{x}-x/\hat{x}\right) ^{2}\prod _{i=1}^{n}f_{k_{i}}
\left( x\hat{x},x/\hat{x}\right) =\left( \int _{0}^{e^{-i
\varepsilon }}\hspace{-1ex} d\hat{x}-\int _{0}^{e^{i
\varepsilon }}\hspace{-1ex} d\hat{x}\right) \frac{\left( x\hat
{x}-x/\hat{x}\right) ^{2}}{2\pi i\hat{x}} \prod _{i=1}^{n}f_{k_{
i}}\left( x\hat{x},x/\hat{x}\right) \nonumber \\ 
&&\hspace{-5mm}=\int _{0}^{1}\frac{d\hat{x}}{2\pi i\hat{x}} 
\left( x\hat{x}-x/\hat{x}\right) ^{2}\left( \prod _{i=1}^{n}f_{k
_{i}}\!\left( x\hat{x}e^{-i\varepsilon },\left( x/\hat{x}
\right) e^{i\varepsilon }\right) -\prod _{i=1}^{n}f_{k_{i}}\!
\left( x\hat{x}e^{i\varepsilon },\left( x/\hat{x}\right) e^{-i
\varepsilon }\right) \right) \nonumber \\ 
&&\hspace{-5mm}=-\int _{0}^{1}\frac{d\hat{x}}{2\pi i\hat{x}} 
\left( x\hat{x}-x/\hat{x}\right) ^{2}\sum _{\ell =1}^{n}\prod _{
i<\ell }f_{k_{i}}\!\left( x\hat{x}e^{-i\varepsilon },\left( x/
\hat{x}\right) e^{i\varepsilon }\right) \nonumber \\ 
&&\cdot \hspace{ 1ex}\left( f_{k_{\ell }}\!\left( x\hat{x}e^{i
\varepsilon },\left( x/\hat{x}\right) e^{-i\varepsilon }
\right) -f_{k_{\ell }}\!\left( x\hat{x}e^{-i\varepsilon },
\left( x/\hat{x}\right) e^{i\varepsilon }\right) \right) \prod _{
i>\ell }f_{k_{i}}\!\left( x\hat{x}e^{i\varepsilon },\left( x/
\hat{x}\right) e^{-i\varepsilon }\right) \nonumber \\ 
\label{8}&&\hspace{-5mm}=-\int _{0}^{x_<}\frac{d\hat{x}}{\hat
{x}} \sum _{\ell =1}^{n}\prod _{i<\ell }f_{k_{i}}\!\left( x\hat
{x}e^{-i\varepsilon },\left( x/\hat{x}\right) e^{i\varepsilon }
\right) \frac{\delta _{\ell }+ \bar{\delta }_{\ell }}{\left( 4
\delta _{\ell }\bar{\delta }_{\ell }\right) ^{k_{\ell }}{}} 
\left( x/|\hat{x}|-x|\hat{x}|\right) \left( x|\hat{x}|\right) ^
{\delta _{\ell }}\left( x/|\hat{x}|\right) ^{-\bar{\delta }_{
\ell }}\\ 
&&\cdot \hspace{ 1ex}\prod _{i>\ell }f_{k_{i}}\!\left( x\hat{x}e^
{i\varepsilon },\left( x/\hat{x}\right) e^{-i\varepsilon }
\right) \nonumber 
\end{eqnarray} 
where $ \prod _{i<1}\equiv \prod _{i>n}\equiv 1$.

\pagebreak[3]

\noindent {\bf Proposition} 8.\  
\begin{equation}
\int _{0}^{\infty }dx\;x\int _{0}^{x_<}\frac{d\hat{x}}{\hat{x}} F
\left( x\hat{x},x/\hat{x}\right) =\frac{1}{2}\int _{0}^{1}dx
\int _{1}^{\infty }\hspace{-1ex} d\bar{x}\;F\left( x,\bar{x}
\right) \hspace{.6ex}.
\end{equation}

\pagebreak[3]

\noindent {\bf Proof}. 
\begin{eqnarray*}
&&\int _{0}^{1}dx\;x\int _{0}^{x}\frac{d\hat{x}}{\hat{x}} F
\left( x\hat{x},x/\hat{x}\right) +\int _{1}^{\infty }dx\;x\int _{
0}^{1/x}\frac{d\hat{x}}{\hat{x}} F\left( x\hat{x},x/\hat{x}
\right) \\ 
&=&\int _{0}^{1}dx\;x\int _{0}^{1}\frac{d\hat{x}}{\hat{x}} F
\left( x^{2}{} \hat{x},1/\hat{x}\right) +\int _{1}^{\infty }dx
\;x\int _{0}^{1}\frac{d\hat{x}}{\hat{x}} F\left( \hat{x},x^{2}
{} /\hat{x}\right) \\ 
&=&\frac{1}{2}\int _{0}^{1}dx\int _{1}^{\infty }\frac{d\hat{x}}{
\hat{x}} F\left( x/\hat{x},\hat{x}\right) +\frac{1}{2}\int _{1}^{
\infty }dx\int _{0}^{1}\frac{d\hat{x}}{\hat{x}} F\left( \hat{x}
,x/\hat{x}\right) \\ 
&=&\frac{1}{2}\left( \int _{1}^{\infty }d\bar{x}\int _{0}^{1/
\bar{x}}dx+\int _{0}^{1}dx\int _{1/x}^{\infty }d\bar{x}\right) F
\left( x,\bar{x}\right) \\ 
&=&\frac{1}{2}\int _{0}^{1}dx\int _{1}^{\infty }d\bar{x}\left( 
\Theta \left( 1/\bar{x}-x\right) +\Theta \left( \bar{x}-1/x
\right) \right) F\left( x,\bar{x}\right) =\frac{1}{2}\int _{0}^{
1}dx\int _{1}^{\infty }\hspace{-1ex} d\bar{x}\;F\left( x,\bar
{x}\right) \hspace{.6ex}.
\end{eqnarray*}\hfill $\Box $
\pagebreak[3] 

\noindent{}With the shorthand 
\begin{equation}
g^{\pm }_{k}\left( x,\bar{x}\right) =\frac{\delta + \bar
{\delta }}{\left( 4\delta \bar{\delta }\right) ^{k}{}} \int _{0
}^{e^{\pm i\varepsilon }}\hspace{-1ex} dt\;t^\delta \left( 
\frac{t}{x\bar{x}}\right) ^{\bar{\delta }}\left( \frac{1}{t-x} 
-\frac{1}{t-\bar{x}}\right) 
\end{equation} 
we get from Eqs.\ (\ref{4}), (\ref{8}) (for $ -1<\alpha <n-3$ all 
integrals converge) 
\begin{equation}
\label{7}B_\alpha \left( {\bf k}\right) =\frac{1}{8}\sum _{\ell 
=1}^{n}\frac{\delta _{\ell }+ \bar{\delta }_{\ell }}{\left( 4
\delta _{\ell }\bar{\delta }_{\ell }\right) ^{k_{\ell }}{}} 
\int _{0}^{1}dx\int _{1}^{\infty }d\bar{x}\frac{x^{\alpha +
\delta _{\ell }}\bar{x}^{\alpha -\bar{\delta }_{\ell }}}{
\left( \bar{x}-x\right) ^{n-2}{}} \prod _{i<\ell }g^{-}_{k_{i}}
\left( x,\bar{x}\right) \prod _{i>\ell }g^+_{k_{i}}\left( x,
\bar{x}\right) \hspace{.6ex}.
\end{equation}

\pagebreak[3]

\noindent {\bf Proposition} 9.\  
For $ x<1$ and $ \bar{x}>1$ 
\begin{equation}
g^{\pm }_{k}\left( x,\bar{x}\right) =\frac{\delta + \bar
{\delta }}{\left( 4\delta \bar{\delta }\right) ^{k}{}} \left( 
\sum _{n=1}^{\infty }\frac{x^{-\bar{\delta }}\bar{x}^{-n-\bar
{\delta }}+x^{n+\delta }\bar{x}^\delta }{n+\delta + \bar
{\delta }}+\pi x^{-\bar{\delta }}\bar{x}^\delta \left( -\cot
\left( \pi \left( \delta + \bar{\delta }\right) \right) \pm i
\right) \right) \hspace{.6ex}.
\end{equation}

\pagebreak[3]

\noindent {\bf Proof}.

\noindent{}(i) 
\begin{displaymath}
\int _{0}^{e^{\pm i\varepsilon }}\hspace{-1ex} dt\;t^\delta 
\left( \frac{t}{x\bar{x}}\right) ^{\bar{\delta }}\frac{1}{t-
\bar{x}} =-\sum _{n=1}^{\infty }\int _{0}^{1}\hspace{-1ex} dt\;t^
\delta \left( \frac{t}{x\bar{x}}\right) ^{\bar{\delta }}\frac{t^{
n-1}}{\bar{x}^{n}{}} =-\sum _{n=1}^{\infty }x^{-\bar{\delta }}
\bar{x}^{-n-\bar{\delta }}\frac{1}{n+\delta + \bar{\delta }} 
\hspace{.6ex}.
\end{displaymath} 
(ii) 
\begin{eqnarray*}
&&\int _{0}^{e^{\pm i\varepsilon }}\hspace{-1ex} dt\;t^\delta 
\left( \frac{t}{x\bar{x}}\right) ^{\bar{\delta }}\frac{1}{t-x} 
=\left( \int _{0}^{\infty e^{\pm i\varepsilon }}\hspace{-1ex} dt
-\int _{1}^{\infty }\hspace{-1ex} dt\right) t^\delta \left( 
\frac{t}{x\bar{x}}\right) ^{\bar{\delta }}\frac{1}{t-x}\\ 
&=&x^\delta \bar{x}^{-\bar{\delta }}\int _{0}^{\infty e^{\pm i
\varepsilon }}\hspace{-1ex} dt\;t^{\delta + \bar{\delta }}
\frac{1}{t-1} -\sum _{n=0}^{\infty }\int _{1}^{\infty }\hspace
{-1ex} dt\;t^\delta \left( \frac{t}{x\bar{x}}\right) ^{\bar
{\delta }}\frac{x^{n}}{t^{n+1}{}} \hspace{.6ex}.
\end{eqnarray*} 
Now, ($ C_{x}^{-1}$ is a negatively orientated circle around $x$) 
\begin{displaymath}
\left( \int _{0}^{\infty e^{+i\varepsilon }}\hspace{-1ex} dt-
\int _{0}^{\infty e^{-i\varepsilon }}\hspace{-1ex} dt\right) t^
{\delta + \bar{\delta }}\frac{1}{t-1} =\int _{C_{x}^{-1}}
\hspace{-1ex} dt\;t^{\delta + \bar{\delta }}\frac{1}{t-1} =-2
\pi i\hspace{.6ex},
\end{displaymath} 
and assuming (without restriction) $ \delta + \bar{\delta }<0$ 
\begin{eqnarray*}
\hspace{-5mm}&&\left( \int _{0}^{\infty e^{+i\varepsilon }}
\hspace{-1ex} dt+\int _{0}^{\infty e^{-i\varepsilon }}\hspace
{-1ex} dt\right) t^{\delta + \bar{\delta }}\frac{1}{t-1} 
\hspace*{1ex}=\hspace*{1ex}2{\rm\hspace{.38ex}Re}\left( -\int _{
0}^{e^{i\varepsilon }}\hspace{-1ex} dt\;t^{\delta + \bar
{\delta }}\sum _{n=1}^{\infty }t^{n-1}+\int _{e^{i\varepsilon }
}^{\infty }\hspace{-1ex} dt\;t^{\delta + \bar{\delta }}\sum _{n
=-\infty }^{0}t^{n-1}\right) \\ 
\hspace{-5mm}&&\hspace*{1ex}=-2\sum _{n=-\infty }^{\infty }{} 
\frac{1}{n+\delta + \bar{\delta }}\hspace*{1ex}=\hspace*{1ex}-2
\pi \cot\left( \pi \left( \delta + \bar{\delta }\right) 
\right) \hspace{.6ex}.
\end{eqnarray*} 
Hence, 
\begin{displaymath}
\int _{0}^{e^{\pm i\varepsilon }}\hspace{-1ex} dt\;t^\delta 
\left( \frac{t}{x\bar{x}}\right) ^{\bar{\delta }}\frac{1}{t-x} 
=-\sum _{n=0}^{\infty }\frac{x^{n-\bar{\delta }}\bar{x}^{-\bar
{\delta }}}{n-\delta -\bar{\delta }} -\pi x^\delta \bar{x}^{-
\bar{\delta }}\left( \cot\left( \pi \left( \delta + \bar
{\delta }\right) \right) \pm i\right) \hspace{.6ex}.
\end{displaymath} 
The $ n=0$-term drops for $ k>0$ by definition of the $\delta $-symbol. 
Changing $\delta $ to $ -\bar{\delta }$ and $ \bar{\delta }$ to 
$ -\delta $ in (ii) gives the result.\hfill $\Box $
\pagebreak[3] 

With $ (\bar{x}-x)^{-n+2}=\sum _{n_{\ell }=1}^{\infty }{n_{\ell 
}+n-4\choose n-3}x^{n_{\ell }-1}\bar{x}^{-n_{\ell }-n+3}$ the $x$-, 
$ \bar{x}$-integrals in Eq.\ (\ref{7}) are trivially evaluated and 
we obtain the following result.

\pagebreak[3]

\noindent {\bf Theorem} 10.\  
Let $ -1<\alpha <n-3$ and $ k_{i}\ge 1$ $ \forall i$. With $ 
\bar{\alpha }=n-4-\alpha $ and 
\begin{equation}
{\cal O}_{j}^{\pm }F\left( \beta _{j},\bar{\beta }_{j}\right) =
\sum _{n_{j}=1}^{\infty }\frac{F\left( -\bar{\delta }_{j},n_{j}
+ \bar{\delta }_{j}\right) +F\left( n_{j}+\delta _{j},-\delta _{
j}\right) }{n_{j}+\delta _{j}+ \bar{\delta }_{j}}+\pi F\left( -
\bar{\delta }_{j},-\delta _{j}\right) \left( -\cot\left( \pi 
\left( \delta _{j}+ \bar{\delta }_{j}\right) \right) \pm i
\right) 
\end{equation} 
we obtain (cf.\ Def.\ 3) 
\begin{equation}
\label{6}B_\alpha \left( {\bf k}\right) =\frac{1}{8}\prod _{i=1}^{
n}\frac{\delta _{i}+ \bar{\delta }_{i}}{\left( 4\delta _{i}\bar
{\delta }_{i}\right) ^{k_{i}}{}} \sum _{\ell =1}^{n}\sum _{n_{
\ell }=1}^{\infty }{n_{\ell }+n-4\choose n-3}\prod _{j<\ell }
{\cal O}_{j}^{-}\prod _{j>\ell }{\cal O}_{j}^+F\left( 
\hbox{\boldmath $\beta$},\overline {\hbox{\boldmath $\beta$}}
\right) \hspace{.6ex},
\end{equation} 
with 
\begin{equation}
F\left( \hbox{\boldmath $\beta$},\overline {
\hbox{\boldmath $\beta$}}\right) =\frac{1}{n_{\ell }+\delta _{
\ell }+\sum _{j\neq \ell }\beta _{j}+\alpha } \;\frac{1}{n_{
\ell }+ \bar{\delta }_{\ell }+\sum _{j\neq \ell }\bar{\beta }_{
j}+ \bar{\alpha }}\hspace{.6ex}.
\end{equation}\hfill $\Box $
\pagebreak[3] 

\vspace{1ex}
\noindent{}In the case $ \alpha =\bar{\alpha }$ there exists a symmetry 
interchanging bared with unbared variables. This allows us to reduce 
the case $ n=4$, $ k_{i}=k$ $ \forall i$ to 6 sums.

With the definitions 
\begin{eqnarray}
A_J&=&\frac{1}{\sum \limits _{j\in J}\!n_{j}+\sum \limits _{j
\in J}\!\delta _{j}-\sum \limits _{j\notin J}\!\bar{\delta }_{j
}+\alpha } \hspace{.6ex},\hspace{2ex}\bar{A}_J\hspace*{1ex}=
\hspace*{1ex}\frac{1}{\sum \limits _{j\in J}\!n_{j}+\sum 
\limits _{j\in J}\!\bar{\delta }_{j}-\sum \limits _{j\notin J}
\!\delta _{j}+ \bar{\alpha }}\hbox{ , for }J\subset \{1,
{\ldots},n\},\nonumber \\ 
C_{j}&=&\frac{1}{n_{j}+\delta _{j}+ \bar{\delta }_{j}}
\end{eqnarray} 
we find 
\begin{eqnarray}
B_{0}\left( k,k,k,k\right) &=&\prod _{i=1}^{n}\frac{\delta _{i}+ 
\bar{\delta }_{i}}{\left( 4\delta _{i}\bar{\delta }_{i}\right) ^{
k}{}} \Big(\sum _{n_{1},n_{2},n_{3},n_{4}=1}^{\infty }\hspace
{-1ex} C_{2}C_{3}C_{4}A_{1234}\bar{A}_{1}+3\hspace{-2ex}\sum _{n
_{1},n_{2},n_{3},n_{4}=1}^{\infty }\hspace{-1ex} C_{2}C_{3}C_{4
}A_{123}\bar{A}_{14}\nonumber \\ 
&&-\;3\pi \cot\left( \pi \left( \delta _{4}+ \bar{\delta }_{4}
\right) \right) \left( \sum _{n_{1},n_{2},n_{3}=1}^{\infty }C_{
2}C_{3}A_{123}\bar{A}_{1}+\sum _{n_{1},n_{2},n_{3}=1}^{\infty }C
_{2}C_{3}A_{12}\bar{A}_{13}\right) \nonumber \\ 
&&+\;\pi ^{2}\left( 3\cot\left( \pi \left( \delta _{3}+ \bar
{\delta }_{3}\right) \right) \cot\left( \pi \left( \delta _{4}+ 
\bar{\delta }_{4}\right) \right) -1\right) \sum _{n_{1},n_{2}=1
}^{\infty }C_{2}A_{12}\bar{A}_{1}{}\\ 
&&-\;\frac{\pi ^{3}}{2}\cot\left( \pi \left( \delta _{2}+ \bar
{\delta }_{2}\right) \right) \left( \cot\left( \pi \left( 
\delta _{3}+ \bar{\delta }_{3}\right) \right) \cot\left( \pi 
\left( \delta _{4}+ \bar{\delta }_{4}\right) \right) -1\right) 
\sum _{n_{1}=1}^{\infty }A_{1}\bar{A}_{1}\Big)\hspace{.6ex}.
\nonumber 
\end{eqnarray} 
Each of the sums is convergent. The last sum is easily evaluated 
to be a multiple of $\pi ^{8}$.

To convert sums 1, 3, 5 into multiple zeta values it is sufficient 
to make repeated use of the identity 
\begin{equation}
\label{5}\frac{1}{n_{1}{}} \frac{1}{n_{2}{}} =\frac{1}{n_{1}+n_{
2}{}} \left( \frac{1}{n_{1}{}} +\frac{1}{n_{2}}\right) 
\end{equation} 
either before of after the expansion of the $\delta $'s.

Sums 2, 4 are first to be converted into partial fractions with respect 
to $k$$_{4}$ or $k$$_{3}$, respectively. Then one uses the identity 
($ i=4,3$) 
\begin{equation}
\sum _{k_{1},k_{i}=1}^{\infty }F\left( k_{i}+k_{1},k_{1}\right) 
=\sum _{k_{1},k_{i}=1}^{\infty }F\left( k_{i},k_{1}\right) -
\sum _{k_{1},k_{i}=1}^{\infty }F\left( k_{i},k_{1}+k_{i}
\right) -\sum _{k_{1}=1}^{\infty }F\left( k_{1},k_{1}\right) 
\end{equation} 
and (\ref{5}) to produce multiple zeta values.

Finally we use tables (due to D.J. Broadhurst) to convert the multiple 
zeta values of weight 8 to a minimum basis consisting of $ 
\zeta (5,3)$, $ \zeta (8)$, $ \zeta (5)\zeta (3)$, and $ \zeta 
(3)^{2}\zeta (2)$. We obtain 
\begin{eqnarray*}
B_{0}\left( 1,1,1,1\right) &=&-\frac{567}{320}\zeta \left( 5,3
\right) +\frac{11663}{3840}\zeta \left( 8\right) -\frac{252}{64
}\zeta \left( 5\right) \zeta \left( 3\right) +\frac{45}{64}
\zeta \left( 3\right) ^{2}\zeta \left( 2\right) \\ 
&&-\frac{351}{320}\zeta \left( 5,3\right) -\frac{1261}{3840}
\zeta \left( 8\right) \hspace*{1ex}+\frac{9}{64}\zeta \left( 5
\right) \zeta \left( 3\right) \hspace*{1ex}+\frac{9}{64}\zeta 
\left( 3\right) ^{2}\zeta \left( 2\right) \\ 
&&\hspace{ 1.7ex}\frac{621}{320}\zeta \left( 5,3\right) -\frac{1
0079}{3840}\zeta \left( 8\right) +\frac{261}{64}\zeta \left( 5
\right) \zeta \left( 3\right) -\frac{90}{64}\zeta \left( 3
\right) ^{2}\zeta \left( 2\right) \\ 
&&\hspace{ 1.7ex}\frac{135}{320}\zeta \left( 5,3\right) +\frac{1
4555}{3840}\zeta \left( 8\right) -\frac{180}{64}\zeta \left( 5
\right) \zeta \left( 3\right) -\frac{18}{64}\zeta \left( 3
\right) ^{2}\zeta \left( 2\right) \\ 
&&-\frac{270}{320}\zeta \left( 5,3\right) -\frac{1710}{3840}
\zeta \left( 8\right) \hspace{ 17.5ex}+\frac{54}{64}\zeta 
\left( 3\right) ^{2}\zeta \left( 2\right) \\ 
&&\hspace{ 13.4ex}-\frac{640}{3840}\zeta \left( 8\right) 
\end{eqnarray*}

\vspace{1ex}
\noindent{}
\begin{equation}
B_{0}\left( 1,1,1,1\right) =-\frac{27}{20}\zeta \left( 5,3
\right) +\frac{261}{80}\zeta \left( 8\right) -\frac{81}{32}
\zeta \left( 5\right) \zeta \left( 3\right) 
\end{equation} 
which reproduces the result in \cite{a} up to a factor of 4$^{5}$ 
which is due to a different normalization of the integrals.

\vspace{1ex}
\noindent{}\section{Conclusions} 
We presented a closed expression for $n$-banana diagrams in terms 
of sums that easily convert to multiple zeta values.

The derivation is valid for $ -1<\alpha <n-3$ whereas the $n$-banana 
is defined for $ -2<\alpha <n-2$. If we approach the limit $ 
\alpha =-1$ or $ \alpha =n-3$ in Eq.\ (\ref{6}) we encounter terms 
of the form $ 1/0^{k}$ for $ k>0$. The limit, however, is finite. 
We conclude that the coefficients in front of the singular terms 
add up to zero. We thus expect Eq.\ (\ref{6}) to be valid in the 
region $ -2<\alpha <n-2$ if one nullifies singular terms.

The inclusion of the case $ k_{i}=0$ for one $ i\in \{0,{\ldots}
,n\}$ is straight forward and amouts to basically replacing the 
numerators $ \delta _{i}+ \bar{\delta }_{i}$ by linear combinations 
of the $ n_{j}$'s. It is hence possible to treat all $n$-banana 
diagrams with our method. Analogous results may be derived in any 
even dimensions.

Equation (\ref{6}) can be used to study the cases where level-mixing 
does not occur. If $ n\ge 5$ we expect in general level-mixing due 
to the inhomogeneous factor of $ {n_{\ell }+n-4\choose n-3}$.

From Eq.\ (\ref{6}) we read off that for $ k_{i}\ge 1$ the 4-banana 
in homogeneous of weight $ 2(k_{1}+k_{2}+k_{3}+k_{4})$. We expect 
the same behaviour if one of the $ k_{i}$'s is zero. This confirms 
the conjecture in \cite{b} that the class $ B_{0}(k,l,1,1)$ of graphs 
related to renormalizable Quantum Field Theories reduces to multiple 
zeta values of weight equal to the crossing number of the related 
knots.

The cases where $ \alpha =-1$, $ \alpha =n-2$, or $ n=3$ are more 
subtle because of the occurence of the singular terms. These cases 
include a second physical series $ B_{0}(m+n+2,k,0)$ which is conjectured 
to lead to weight $ 2(n+m+k)+5$ multiple zeta transcendentals. These 
together with two more classes of non-banana diagrams are expected 
to exhaust all independent multiple zeta values up to weight 16 
(corresponding to 9 loops).

We would not be surprised if future work reveals a simpler derivation 
of $ B_{0}(1,1,1,1)$ since $ B_{0}(1,1,1,1)$ seems genuinly to be 
a two-fold sum: The coefficients in front of the $ \zeta (3)^{2
}\zeta (2)$-transcendental add up to zero. Since the reduction of 
weight 8 depth 2 multiple zeta values (in contrast to depth $ >2
$) never produces $ \zeta (3)^{2}\zeta (2)$-terms the miracle of 
the vanishing coefficient could be naturally avoided if there was 
a calculation of $ B_{0}(1,1,1,1)$ that does not make use of depth 
$ >2$. Moreover our calculation does not explain the connection 
to the (4,3)-torus-knot (and the T$_{5,2}$T$_{3,2}$-factor knot).

We close the paper with the conjecture that any amplitude reduces 
to multiple zeta values if it has an angular graph that reduces 
to a single line ($ \bullet \hspace{-1ex}-\hspace{-1ex}-\hspace
{-1ex}\bullet $) after repeated substitutions of double lines ($ 
\bullet \hspace{-1.3ex}=\hspace{-.5ex}=\hspace{-1.3ex}\bullet $) 
and iterated lines ($ \bullet \hspace{-1ex}-\hspace{-1ex}-
\hspace{-1ex}\bullet \hspace{-1ex}-\hspace{-1ex}-\hspace{-1ex}
\bullet $) by single lines. Those amplitudes are free of 6- or higher-$j$-symbols. 
They do not entail the full structure of 4 dimensions and are therefore 
accessible to reducing 4-dimensional integrals to integrals over 
the complex plane which lies in the heart of our calculations.


\begin{thebibliography}{1}  
\bibitem{a}D.J. Broadhurst, D. Kreimer, {\it Knots and Numbers in 
$\phi ^{4}$ Theory to 7 Loops and Beyond}, hep-ph/9504352. 
\bibitem{b}D.J. Broadhurst, D. Kreimer, {\it Association of multiple 
zeta values with positive knots via Feynman diagrams up to 9 loops}, 
hep-th/9609128. 
\bibitem{c}D. Kreimer, {\it Renormalization and Knot Theory}, q-alg/9607022. 
\bibitem{d}D.J. Broadhurst, J.A. Gracey, D. Kreimer {\it Beyond 
the triangle and uniqueness relations: non-zeta counterterms at 
large N from positive knots}, hep-th/9607174. 
\bibitem{e}D.J. Broadhurst, {\it Solving differential equations for 
3-loop diagrams: relation to hyperbolic geometry and knot theory}, 
hep-th/9806174. 
\end{thebibliography}
\end{document}